\documentclass[
 reprint,
 amsmath,amssymb,
 aps,
]{revtex4-1}
\usepackage{graphicx}% Include figure files
\usepackage{dcolumn}% Align table columns on decimal point
\usepackage{bm}% bold math
\usepackage[version=4]{mhchem}

\begin{document}

\title{Discovery Of Copper Catalyst For New Chemical Polishing Acid \\ On Superconducting Niobium }% Force line breaks with \\

\author{D. Luo}
 \altaffiliation{Luo@frib.msu.edu}%Lines break automatically or can be forced with \\
\author{K. Saito}%

\affiliation{%
 Facility for Rare Isotope Beams, National Superconducting Cyclotron Laboratory, Michigan
State University, East Lansing, MI 48824, USA
}%

\date{\today}% It is always \today, today,

\begin{abstract}
 High field Q-slope (HFQS) seriously limits the high gradient performance of the buffered chemically polished (BCP) superconducting radio frequency (SRF) niobium cavities. The direct cause or mechanism is not yet fully understood. In our recent extensive data analysis, we concluded that the potential root cause could be nitrogen contamination on the surface. The nitric contamination could be created by nitric acid during BCP which uses the mixture of hydrofluoric acid ($\ce{HF}$), nitric acid ($\ce{HNO3}$), and phosphoric acid ($\ce{H3PO4}$). Based on this thought, we started to develop a new chemical polishing acid that replaces the nitric acid by hydrogen peroxide. We have discovered that this new acid cannot provide smooth surface finishing, however adding copper catalyst allows this acid to provide a smooth surface similar to or even better than that from the conventional BCP. This paper first shows the significance of resolving HFQS, then summarizes our extensive data analysis results, and finally describes our discovery of the copper catalyst.
\begin{description}
\item[PACS numbers]
May be entered using the \verb+\pacs{#1}+ command.
\end{description}
\end{abstract}

\pacs{Valid PACS appear here}% PACS, the Physics and Astronomy
\maketitle

%\tableofcontents

\section{Introduction}
\subsection{Demand for new Acid}

For superconducting Nb cavities, chemical polishing or electropolishing (EP) is required to remove defects and the contaminated surface layer. It is an important process that lead to much better performance. One commonly used method -- Buffered Chemical Polishing (BCP) -- always leads to cavity High Field Q-slope (HFQS) which seriously limits the cavity performance at high operating field. The other method -- EP \cite{Kako-improve} -- can recover the cavity HFQS by an extra 120 $^{\circ}$C low temperature baking (LTB) post EP \cite{Lilje}. However, EP is not always applicable to low/medium $\beta$ cavities because of their complicated shapes. Therefore, a new chemical polishing process is in demand, especially for low/medium $\beta$ cavities which are used in heavy ion accelerators.

HFQS is the phenomenon where Q$_0$ (unloaded Q) performance of the SRF cavity begins to drop exponentially when the accelerating gradient increases beyond 80 - 100 mT (corresponding to an accelerating gradient $E_{\text{acc}}$ of 20 - 25 MV/m for ILC elliptical shape cavity \cite{Hasan2chp5}). The Q$_0$ drop is caused by pure heating at RF high magnetic field region (equator area) on the SRF surface \cite{Saito-Qslope,Visentin-BCP,Safa}, and it ultimately limits the accelerating gradient to below 130 mT ($E_{\text{acc}}$ is 30 MV/m for ILC elliptical shape cavity). The pure HFQS is thermal heating and has no X-ray. However, HFQS and X-ray occurrence are mixed in some cases. 

If the HFQS issue is resolved, we can solve one of the most serious performance limitation and increase the accelerating gradient above $E_{\text{acc}}$ = 30 MV/m,   for ILC type elliptical shape cavity \cite{Saito-State}. It can make the accelerator system more compact or increase accelerated particle energy, which will result in a remarkable cost reduction in the SRF project. 

Low to medium $\beta$ cavities evolved in many areas and are becoming one of the most widespread types in LINACs. FRIB is an example of a heavy ion accelerator project whose cavities suffer from HFQS. Nearly 300 cavities (90\% of the total), comprised of 120 low $\beta$ QWRs (quarter wave resonators) and 180 medium $\beta$ half wave resonators (HWRs), have been cold tested. All of these cavities are treated with BCP. Statistically, the performance of $\sim$ 35\% of the cavities at FRIB is limited by pure HFQS (HFQS without X-rays) \cite{Saito-TTC}. If one takes into account the fact that HFQS involves field emission in many case, the probability of a cavity having HFQS $\sim$ 50\% or higher. 

\begin{figure}
\centering
\includegraphics*[width=\columnwidth]{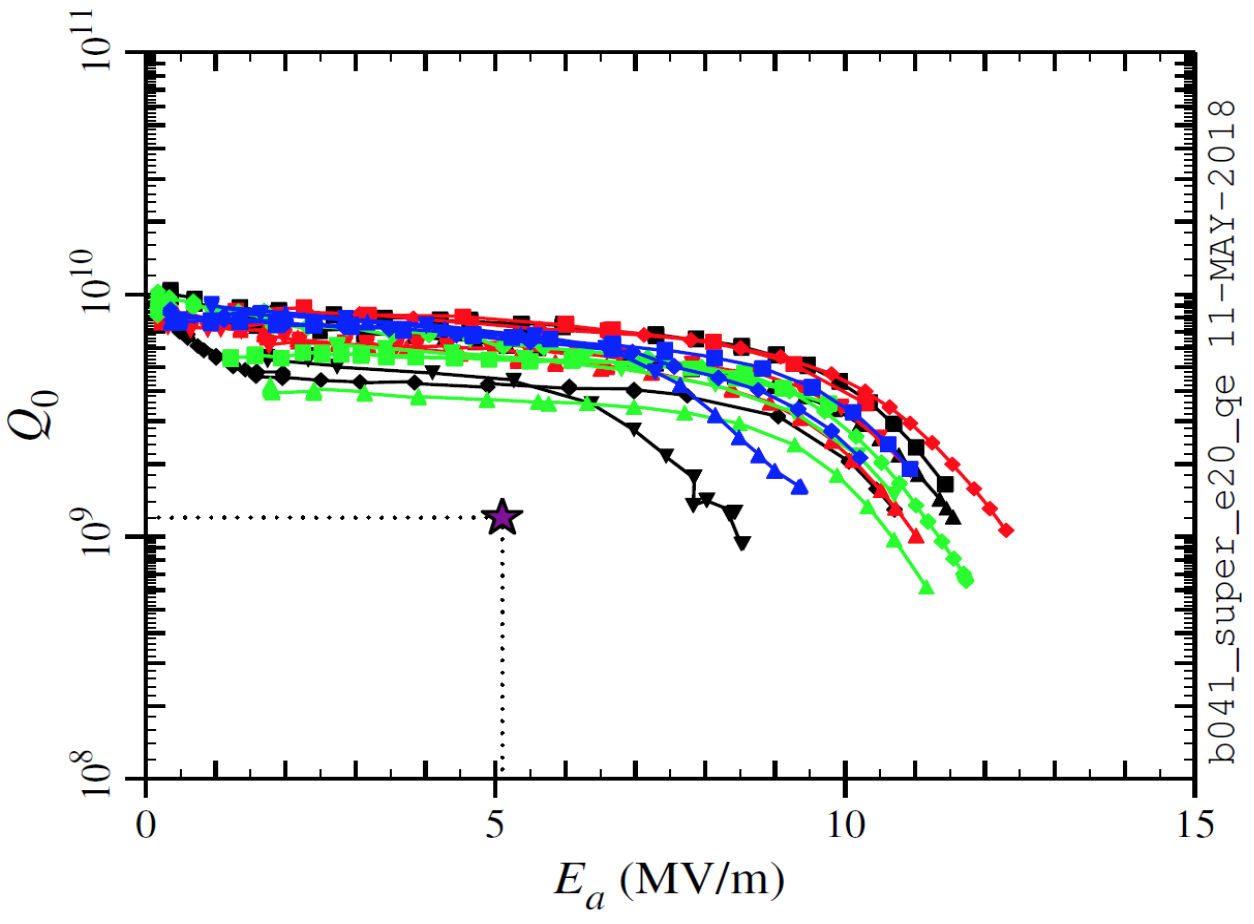}
\caption{FRIB cavity performance at 2 K in Vertical Test, $\beta$ = 0.041 QWRs, $B_{\text{p}}/E_{\text{acc}}$ = 10.71 [mT/(MV/m)]. \label{f:001}}
\end{figure}

In Fig.~\ref{f:001}, the FRIB cavity performance is presented by $Q_0$ vs $E_{\text{acc}}$. Converting $E_{\text{acc}}$ to $B_{\text{p}}$ with the design ratio 10.71 [mT/(MV/m)], the $Q_0$ starts to drop from $B_{\text{p}}$ $\sim$ 85 mT in $\beta$ = 0.041 QWRs. If HFQS is resolved, these cavities have the potential to operate at 10 MV/m (instead of the current gradient 5 MV/m). Similar considerations apply to other cavities in FRIB, which mean the SRF LINAC can almost be shortened by half or the acceleration energy can be doubled.

\subsection{Alternative Acid Treatment}

It is highly probable form our recent analysis that the HFQS under BCP arises from nitrogen contamination, which is introduced by nitric acid in the commonly used BCP acid. Thus, the replacement of the nitric acid by an alternative is a promising way to mitigate this issue. For this purpose, we summarize related experimental results as follows \cite{Luo-2019}:

\begin{enumerate}
    \item If a cavity was vented with nitrogen gas after cold test and exposed for one night, the extra cold test afterward shows the cavity was often limited by Q-slope. Replacing nitrogen gas by argon gas preserved the flat Q up to 40 MV/m \cite{Saito-Basic}.
     \item If nitric acid (1500ppm) is added to EP, subsequent LTB can no longer always eliminate the HFQS \cite{Higuchi}.
    \item The chemical barrel polishing (CB) using BCP + EP + LTB produced a sharp quench at below 20 MV/m on a cavity, while the barrel CB using water + EP + LTB provided high field performance $E_{\text{acc}}$ $>$ 30 MV/m. Microscope observation of the niobium sample by the CB with BCP + EP showed niobium-nitride (Nb-N) like crystal structures (triangular structures) on the surface \cite{Higuchi-p}. 
    \item BCP HFQS has a deep memory effect which cannot be explained only by surface smoothness changes \cite{Kako-improve,Luo-2019}.
    \item Large grain/single crystal cavities have very smooth surface after etching by BCP ($R_{\text{z}}$ $\sim$ 0.2 $\mu$m), but their highest achievable gradient before quenching occurs are still lower than those of EP'ed fine grain cavities (40 MV/m in case of ILC shape). \cite{Kneisel-Large,Kneisel-single,Singer-1,Singer-2}. 
    \item Nitrogen doping technique shows that the Nb-N phase generated on the SRF top surface has very harmful impact on cavity performance. This layer has to be removed by EP $\sim$ 5 $\mu$m \cite{grassellino2017unprecedented}.
     
 \end{enumerate}

 Based on these experiments, we formulated three criteria on the alternative acid:
 
 First, the surface roughness after polishing should be smoother than 3 $\mu$m. It has been reported that high surface roughness will lead to magnetic field enhancement \cite{Knobloch,Xu}, and result in flux trapping \cite{Ben-fluxtrap}. A previous simulation of EP'ed cavities suggests field enhancement will start from a surface roughness of $\sim$ 3 $\mu$m \cite{Saito-Roughness}, that is, the finished surfae roughness should be smoother than 3 $\mu$m. 
 
 Second, material removal speed should be at a level of 2 - 5 $\mu$m/min, similar to conventional BCP. This is important for real cavity  processing. 

 Finally, the alternative acid should be nitrogen-free to prevent nitrogen contamination.
 
 Here, hydrogen peroxide ($\ce{H2O2}$) is the first candidate that comes to our mind for replacing the nitric acid as a new oxidizer, since it does not introduce any extra elements contamination and is itself a strong oxidizer. An extra benefit is that the reaction will not generate any hazardous NO$_x$ gas.

\subsection{Past acid R$\&$D with $\ce{H2O2}$ }

Development of this kind of acid has been already attempted by two institutes. One was in the early 1980's to resolve the issue in waste water treatment with the conventional BCP acid \cite{Bloess}. This mixture was unstable and the polished surface was not smooth and they concluded it was due to the higher water content. The second one was to improve the BCP'ed cavity performance \cite{Antonie} at the end of the 1990's at Saclay, France. They were trying to find alternative chemical treatment of the conventional BCP. Various mixture of $\ce{HF}$ (5 - 15 mol/L) and $\ce{H2O2}$ (2.5 - 5 mol/L) were applied to niobium samples. However, the surface was always degraded by severe grain boundary etching and significant roughness (mean peak-to-valleys $>$ 50 $\mu$m). 

\section{Experimental Method and First Results}

To maintain consistency with previous published data sets, we take $R_{\text{z}}$ as a measure of surface roughness defined by:

\begin{equation}
R_z=\frac{1}{5} \sum_{i=1}^{5}(R_{p,i}-R_{v,i})
    \label{Eqs:Rz}
\end{equation}
where $R_{p,i}$ and $R_{v,i}$ denote the $i^{th}$ highest peak and $i^{th}$ lowest valley within the evaluation length respectively \cite{wiki:xxx}. $R_{\text{z}}$ is directly measured by the stylus (roughness tester).

\subsection{Experimental Setup}
 
Prior to the experiments, niobium samples (50 mm $\times$ 15 mm $\times$ 4 mm) were mechanically polished by emery paper to adjust the initial surface roughness. The surface roughness after etching was measured at no fewer than 5 points near the initial measurement points and the results were averaged. The standard deviation was used to construct the error bar. The information on the initial surface roughness is shown in Table~\ref{tab:table1}.

\begin{table}
\caption{\label{tab:table1}
Mechanical polishing prior to experiment and the initial surface roughness}
\begin{ruledtabular}
\begin{tabular}{lcc}
    &  Mechanical polishing   & Initial   \\
     & with emery paper
 & roughness\\
 & (number of roughness)& $R_{\text{z}}$ [$\mu$m]

\\
 \colrule
Baseline BCP         & $\#$ 320 & 4.6 \\
First attempt        & $\#$ 320 & 4.3\\
First optimization   & $\#$ 320 & 4.6\\
Second optimization  & $\#$ 320 & 3.7/2.2\footnote{2.2 $\mu$m is for the 409 ppm case, and it adopts an improved way to mechanically polish the surface.}\\
Third optimization   & $\#$ 600 & 2.4
\end{tabular}
\end{ruledtabular}
\end{table}

\subsection{First Test of the New Acid: 50\% $\ce{HF}$ + 50\% $\ce{H2O2}$}
\begin{figure*}
\centering
\includegraphics*[width=1.5\columnwidth]{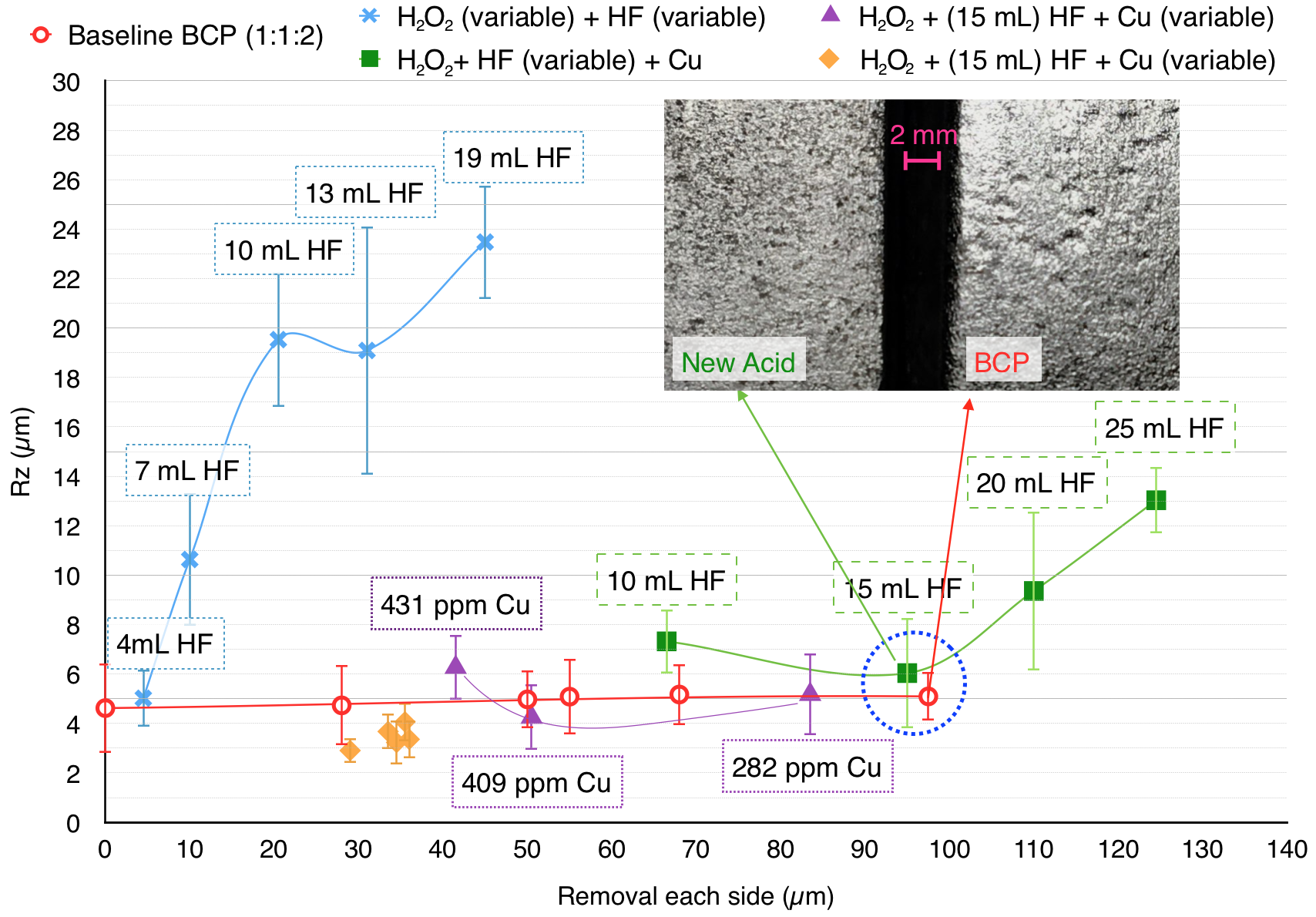}
\caption{Optimization of new acid process and BCP baseline data with material removal vs surface roughness. The inserted picture shows a comparison of samples processed with new acid (left) and BCP (right), the two samples has similar initial roughness ($\sim $4.5 $\mu$m) and similar total removal ($\sim$ 100 $\mu$m). BCP sample is more shiny but has corrosion (orange skin). \label{f:002}}
\end{figure*}

We started the development of this acid mixture again. We reconfirmed that this acid provides very rough surface as material removal increases. For instance, for fine grain niobium material with RRR = 250 - 300, this mixture resulted in $R_{\text{z}}$= 20 - 25 $\mu$m as shown in Fig.~\ref{f:002} (blue cross marks), while the conventional BCP acid can attain a roughness of $\sim$ 5 $\mu$m (red empty circle in Fig.~\ref{f:002}). In this experiment, we fixed the total volume of acid at 55 mL, and changed the acid ratio. No optimum point could be obtained When the volume of 50\% $\ce{HF}$ was varied from 4 mL to 19 mL and that of 50\% $\ce{H2O2}$ was varied from 51 mL to 36 mL correspondingly. 24 mL HF + 31 mL $\ce{H2O2}$, 27 mL HF + 28 mL $\ce{H2O2}$ experiments were also done, but the roughness were beyond the roughness tester's measurement range (25.2 $\mu$m). The acid bath temperature was not actively controlled, the initial temperature was 18 $^{\circ}$C and the final temperature ranged from 35 to 60 $^{\circ}$C. This is the same in other experiments.

Large grain niobium samples were used to investigate the reason. The samples were composed of two large crystalline with sizes $\sim$ 3 cm $\times$ 1 cm. These two crystalline had different crystal orientation. We observed a big difference in surface roughness after $\sim$ 20 $\mu$m etched in these crystalline. One had $R_{\text{z}}$ = 6.8 $\pm$ 2.5 $\mu$m and another had $R_{\text{z}}$ = 16.6 $\pm$ 3.6 $\mu$m. This provided evidence to suggest that preferential etching depends on the crystal orientation.

\subsection{Other Trials}
The viscosity of etching acid can influence the performance of BCP as follows. It is effective to slow the diffusion electrolyte reaction products by the chemical reaction, making the morphology: peak area has higher acid concentration, and vice versa for the valley area. In other words, it prevents the preferential etching. This effect is enhanced when the viscosity is high \cite{shigolev, Viscosity, Viscosity2}.

In order to study the viscosity effect, we then tried $\ce{HF}+\ce{H2O2}+\ce{H3PO4}$ mixture because phosphoric acid has high viscosity and is commonly used in BCP. However, the results showed that only material removal speed was reduced while the surface roughness remained very rough. 

$\ce{HF}+\ce{H2O2}+\ce{H2SO4}$ mixture was used in the second trial. The reason of the material speed reduction in the above mixture is probably due to the poor oxidation capability. Sulfuric acid will add oxidation power as well as high viscosity to the new acid. We attempted this mixture. The results showed surface roughness $R_{\text{z}}$ = 10.4 $\pm$ 6.2 $\mu$m after $\sim$ 8.4 $\mu$m etched. The mixture consisted of 10 mL 50\% $\ce{HF}$, 45 mL 35\% $\ce{H2O2}$, and 60 mL 98\% $\ce{H2SO4}$. We also tried other composition ratios, but the results had no significant differences. This mixture will continue to be optimized, but it is expected to be a complex optimization.

Several other parameters were also investigated: component concentration, $\ce{H2O2}$ decomposition, initial temperature, and agitation effect. However, none of their variation reduced the roughness. Based on our results: 1) initial temperature and viscosity variation only affects the reaction rate; 2) adding agitation makes the surface rougher; and 3) non-stabilized $\ce{H2O2}$ experiment shows that decomposition has no obvious effect.

\subsection{New Attempt}
Before abandoning this series of acid mixture, we applied this mixture on Cu metal to observe whether the same problems would occur. To our surprise, the result was a very smooth finishing surface. This led us to attempt adding copper in the acid.

\section{Copper Catalyst -- Towards The Optimization}

We made several tests and found that the method works: acid reaction with a small amount Cu prior to immersing Nb can lead to very different results. We also found high concentration 50\% $\ce{H2O2}$ generates huge bubble marks on the Nb surface, so we added DI water to adjust $\ce{H2O2}$ to 35\% for a standard start.

\subsection{First Attempt}
In the first attempt, we used an acid mixture of total volume 55 mL, which consisted of 10 mL 50\% $\ce{HF}$, 29 mL 50\% $\ce{H2O2}$, 16 mL $\ce{H2O}$ and 16 mg copper powder. The acid bath temperature was not actively controlled, its initial temperature was 18 $^{\circ}$C and the final temperature varied from 35 to 45 $^{\circ}$C. Material removal speed was $\sim$ 4$\mu$m/min. The finished surface roughness reached $R_{\text{z}}$ = 7 $\mu$m after 65 $\mu$m etched as shown in Fig.~\ref{f:002} (solid square mark). This result was much better than almost all the cases that were done without Cu.
 
 The new acid generates no NO$_x$ gas and the finishing surface has no orange skin. This is a breakthrough for our new acid development. Then we started optimizing the acid with the control parameter method.

\subsection{First Optimization: HF Concentration Dependence}
Building upon the first attempt, optimization of $\ce{HF}$ concentration was conducted (Fig.~\ref{f:002}, solid green square marks). We first fixed the total amount of acid mixture (55 mL), the amount of 29 mL 50\% $\ce{H2O2}$, and the amount of Cu (300 ppm). The concentration of $\ce{HF}$ was adjusted to 26 mL by adding water to the 50\% $\ce{HF}$. For instance in Fig.~\ref{f:002} (green solid square marks), 15 mL $\ce{HF}$ means 15 mL of 50\% $\ce{HF}$ acid and 11 mL of water. The reaction speed increased rapidly with $\ce{HF}$ concentration, and so did the finishing surface roughness. We found the $\ce{HF}$ amount at 15 mL gave the best surface roughness, and we chose it for subsequent experiments shown by the solid triangles in Fig.~\ref{f:002}. 

\subsection{Second Optimization: Cu Amount Dependence}
We fixed all other parameters and only changed the amount of Cu. We found that 409 ppm Cu gave the best result. In this case, we adjusted the initial roughness to around 2 $\mu$m to study the impact of the initial surface. For the first time, we achieved better surface roughness than conventional BCP. 

\subsection{Third Optimization: Accurate Control of Cu and Initial Roughness}
 Previous copper optimization was done by copper powder, which posed difficulties in weigh control because electrostatic force made the powder easily spill out of the container. We repeated copper optimization using copper wire, and increased the accuracy of copper concentration. In this experiment, we polished niobium samples mechanically with emery paper of $\#$600 to get a more standard initial surface roughness. The material removal speed is $\sim$ 4 $\mu$m/min.
 
\begin{figure}
 \centering
 \includegraphics*[width=\columnwidth]{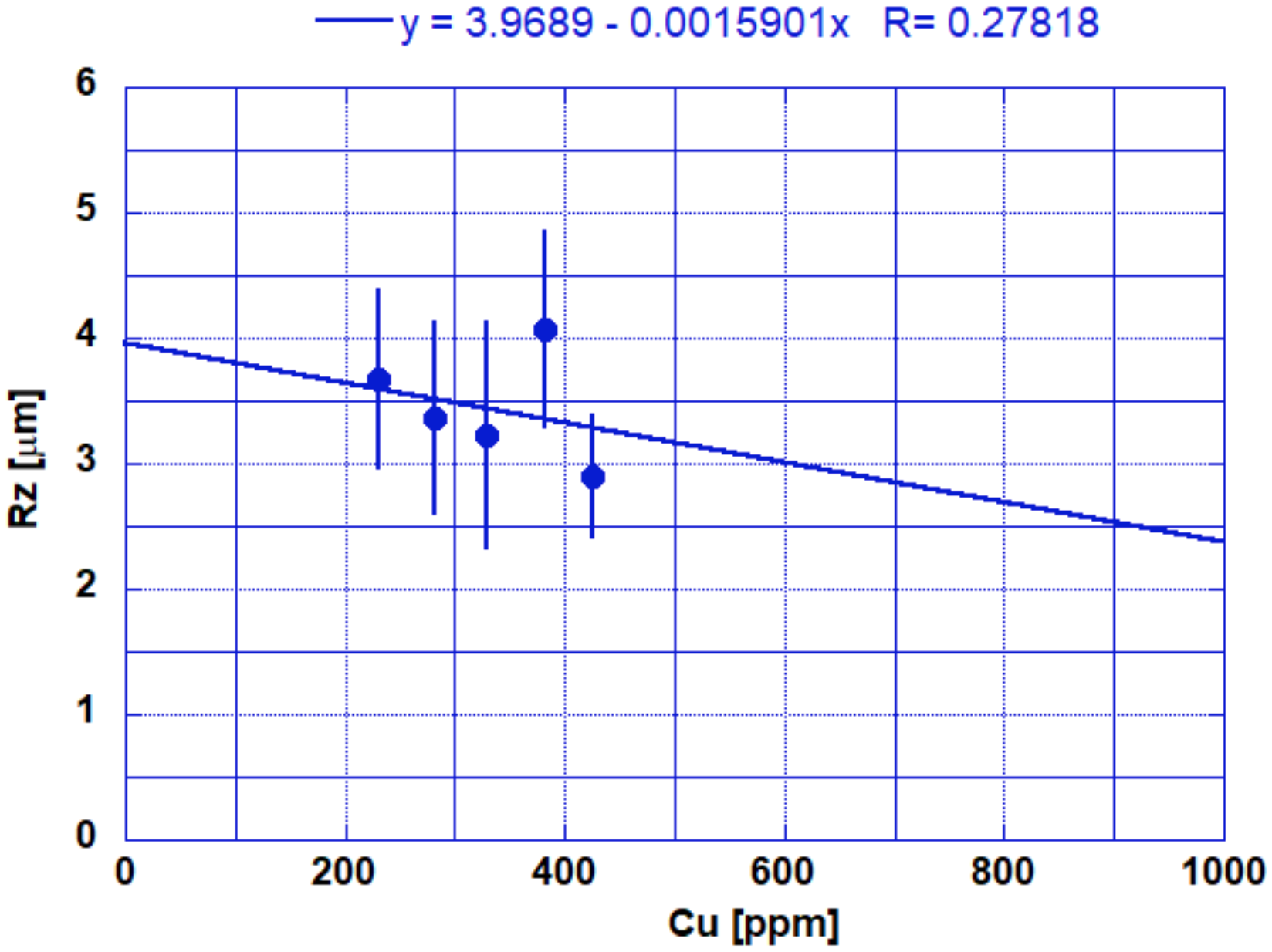}
 \caption{ Third optimization results zoomed in Fig.~\ref{f:002} with Cu amount as the x axis. A lower than 2.5 $\mu$m finishing surface roughness is expected at the copper concentration of 900 ppm with linear fitting. \label{f:003}}
 \end{figure}

 This effect is shown in Fig.~\ref{f:002} (orange diamond marks) and Fig.~\ref{f:003} (with Cu amount as x axis). In Fig.~\ref{f:003}, based on the results with copper concentration in the range of 250 - 425 ppm, a finishing roughness of 3 $\mu$m can be achieved. A linear extrapolation shows that 2.5 $\mu$m seems possible at 900 ppm. More studies will be done to reduce errors and better establish the dependence.
 
We also confirmed that the finishing surface roughness depends on the initial roughness because all cases in the third optimization had lower finishing roughness than before, no matter how the Cu amount varied. A smoother initial roughness produces a smoother finishing surface. This means that this new acid has no crystal orientation preferential etching.

\section{Mechanism}

This section explains the chemical reactions that occur when Cu is added into the polishing acid mixture and discusses their implications.

A series of reactions happen during the experiment: 
\begin{equation}
    \ce{Cu + H2O2 -> CuO + H2O}
    \label{equation:001}
\end{equation}
\begin{equation}
    \ce{CuO + H2O2 -> CuO2 + H2O}
    \label{equation:002}
\end{equation}
\begin{equation}
    \ce{2 CuO2 -> 2 CuO + O2 ^}
    \label{equation:003}
\end{equation}
These reactions constitute a $\ce{H2O2}$ decomposing process with Cu as the catalyst, and can be written as: 
\begin{equation}
    \ce{2 H2O2 ->[Cu] 2 H2O + O2 ^}
    \label{equation:004}
\end{equation}
Then the \ce{CuO} reacts with $\ce{HF}$:  
\begin{equation}
    \ce{CuO + 2 HF -> CuF2 + H2O}
    \label{equation:005}
\end{equation}
Or
\begin{equation}
   \ce{CuO + 4 HF -> Cu(HF2)2 + H2O}
   \label{equation:006}
\end{equation}
The $\ce{Cu^2+}$ dissolves in the acid, and when Nb is put into the acid, reaction
\begin{equation}
   \ce{5 Cu^2+ + 2 Nb -> 5 Cu + 2 Nb^5+}
   \label{equation:007}
\end{equation}
happens simply due to single-displacement reaction, and $\ce{Nb^5+}$ dissolves into the acid to generate a Cu layer outside the Nb, and reaction (1) to (7) continuously happen, until any of the component is totally consumed. During this process, Cu accelerates the reaction and is not consumed, thus acting as catalyst.

The combination of the whole process can be written as:
\begin{equation}
    \ce{20 H2O2 + 20 HF + 4 Nb ->[Cu]  4 NbF5 + 5 O2 ^ + 30 H2O}
    \label{equation:008}
\end{equation}

The key components of the whole reaction series are reaction (4) and reaction (7): there will be no gas product without reaction (4), which is important to finishing surface roughness (will be explained later); if there is no reaction (7), the metal ions will be excluded from the reaction series and just stay in the acid. Therefore, any metals that meet these two processes should be able to work as a catalyst for this acid. Reaction (7) needs the metal to be less reactive (in the reactivity series) than Nb, and reaction (2) needs the metal to be at least oxidizable by $\ce{H2O2}$ (can be found from Standard electrode potential), so the series of potential catalysts is: (Nb), Zn, Cr, Ga, Fe, Cd, In, Tl, Co, Ni, Mo, Sn, Pb, W, Ge, Cu, Tc, Ru, Po, Hg, Ag.

In this series, the more reactive (closer to Nb) the metal is, the slower reaction (7) will be, which is closer to the scenario without catalyst. Conversely, the less reactive (closer to Ag) the metal is, the stronger the metal ion's oxidization power, so it is more catalyst effective.

We also tried Fe, which is closer to Nb than Cu, and the finishing surface roughness was $R_{\text{z}}$ = 10 $\sim$ 16 $\mu$m depending on the amount of Fe. 

\section{Discussion}

The underlying principle of general chemical polishing is a higher reaction rate at the peaks than at the valleys. This produces a smoothing effect on the surface, some fundamental studies can be found in Ref.~\cite{shigolev}. Several parameters are important for the finishing surface, such as viscosity \cite{Viscosity, Viscosity2}, oxidation layer formation \cite{Cathode}, temperature etc..

For a chemical polishing process, the reaction rate difference arises because the peak has higher acid amount to react with than the valley for a similar reaction area \cite{shigolev, Peak-valley}. This difference can be enhanced by bubbles generated on the metal surface which block the reaction more around the valley area than at the peak \cite{Guo}. Bubbles that grow during the reaction will start departing the surface once they reach a critical condition, which is more easily met at the peak than at the valley (see Ref.~\cite{bubble1, bubble2PhysRevFluids.3.073602}). Hence, bubbles take a longer time to depart from valleys and block the reaction at valleys for a longer time than they do at peaks. This effect enlarges the reaction rate difference and facilitates the generation of a smoother surface.

The reaction of the $\ce{H2O2}$ plus $\ce{HF}$ with Nb is: 
\begin{equation}
    \ce{5 H2O2 + 2 Nb + 10 HF -> 2 NbF5*H2O + 8 H2O}
    \label{equation:009}
\end{equation}
One obvious difference between this reaction (Eqs.~\ref{equation:009}) and the Cu case (Eqs.~\ref{equation:008}) is the $\ce{O2}$ gas generated on Nb surface, and it is possible that these oxygen gas bubbles help generate smooth surface.

There is an extra benefit from adding Cu: the Cu generated on Nb surface from Eqs.~\ref{equation:007} forms cathode, increases the total reaction rate (especially on the peaks because the peaks already react faster \cite{Peak-valley}) and intensifies the reaction rate difference. This is due to Galvanic corrosion, which is similar to the aluminum case in Ref.~\cite{Cathode}.

\begin{acknowledgments}

This work was supported by the U.S. National Science Foundation under Grant PHY-1565546; and by Michigan State University. The authors are grateful to J. Taguchi and Y. Okii in Nomura Plating Co. Ltd., Japan; and Ethan Metzgar, Laura Popielarski, Safwan Shanab, George Vernon Simpson, Yoshishige Yamazaki and Jie Wei from FRIB for great help with sample surface treatments and valuable discussions. 

\end{acknowledgments}

\bibliography{Luo_06142019}

\end{document}